\documentclass[12pt]{article}
\usepackage{epsfig}
\usepackage{amsmath}
\usepackage{amssymb}

\date{\today}

\newcommand{\be}{\begin{equation}}
\newcommand{\ee}{\end{equation}}
\newcommand{\bea}{\begin{eqnarray}}
\newcommand{\eea}{\end{eqnarray}}
\newcommand{\bml}{\begin{mathletters}}
\newcommand{\eml}{\end{mathletters}}

\begin{document}
\thispagestyle{empty}
\vspace*{1.cm}
\begin{center}
{\Large \bf 
How fast can a black hole rotate?
}
\\
\vspace{1.5cm}
 {\large Carlos A. R. Herdeiro\footnote{herdeiro@ua.pt} and Eugen Radu\footnote{eugen.radu@ua.pt}} 
\\
\vspace{0.5cm}
\small{
 Departamento de F\'\i sica da Universidade de Aveiro and CIDMA, 
   \\
   Campus de Santiago, 3810-183 Aveiro, Portugal} 
\end{center}

\begin{abstract} 
Kerr black holes have their angular momentum, $J$, bounded by their mass, $M$: $Jc\leqslant GM^2$. There are, however, known black hole solutions violating this \textit{Kerr bound}. We propose a very simple universal bound on the rotation, rather than on the angular momentum, of four-dimensional, stationary and axisymmetric, asymptotically flat black holes, given in terms of an appropriately defined \textit{horizon linear velocity}, $v_H$. The $v_H$ bound is simply that $v_H$ cannot exceed the velocity of light. We verify the $v_H$ bound for known black hole solutions, including some that violate the Kerr bound, and conjecture that only extremal Kerr black holes saturate the $v_H$ bound. 

\end{abstract}

\vspace*{1.cm}

\begin{center}
{\footnotesize{Essay written for the Gravity Research Foundation 2015 Awards for Essays on Gravitation.}}
\\
{\footnotesize{Submitted March 15th, 2015}}

\end{center}
\newpage
\vspace{0.5cm}


\section{Motivation: how strict is the Kerr bound\,?}
\label{section1}

Macroscopic objects from our daily experience can have a total angular momentum, $J$, larger than their total mass, $M$, squared:
\begin{equation}
\frac{Jc}{GM^2}\equiv j >1 \ .
\end{equation} 
In fact, in view of the SI values of Newton's constant $G\simeq 6.67\times 10^{-11}$ and of the speed of light, $c\simeq 3\times 10^{8}$, this is inevitable for objects with $J,M\sim 1$. For such objects -- for instance, a spinning football -- $j$ is many orders of magnitude above unity. In essence, this is because electromagnetic interactions, rather than gravity, keep the object's structure. 

By contrast, self-gravitating \textit{compact} objects, for which the total energy is dominated by gravitational binding energy,  should have $j \lesssim 1$. Indeed, considering one such object with radius $R$, and requiring that the magnitude of the gravitational binding energy $|E_{\rm gra}|\sim GM^2/R$ should not be smaller than the object's rotational energy $E_{\rm rot}\sim J^2/(MR^2)$, implies that $j \lesssim \sqrt{{GM}/(c^2R)}$.
%
Thus, if the system has a dimension of the order of its gravitational radius $R\sim GM/c^2$, then $j \lesssim 1$. This agrees with what is inferred from the most general vacuum, stationary, asymptotically flat, four-dimensional black hole (BH) solution of general relativity, the Kerr metric~\cite{Kerr:1963ud}. This solution obeys the \textit{Kerr bound}, $j \leqslant 1$. As argued in the Misner-Thorne-Wheeler treatise~\cite{Misner:1974qy} ``it seems likely that in any collapsing body which violates this constraint, centrifugal forces (...) will halt the collapse before a size $\sim GM/c^2$ is reached". 

The above heuristic arguments make plausible that any physically reasonable BH solution should obey $j\lesssim 1$ as a ballpark estimate.
 For the Kerr BH the bound is actually \textit{strict}, $j \leqslant 1$, a result that has been extended to vacuum, \textit{dynamical}, axi-symmetric and asymptotically flat BHs~\cite{Dain:2005vt,Dain:2006wb,Chrusciel:2007ak}. Is, thus, this strict Kerr bound a fundamental relation for \textit{any} four dimensional, asymptotically flat and physically reasonable BH?

\newpage

\section{Black holes that violate the Kerr bound}
To answer the above question it suffices to exhibit a BH solution violating the Kerr bound.\footnote{Throughout this essay we focus on four dimensional, stationary, regular,
asymptotically flat
 BH solutions with physically acceptable matter, unless otherwise explicitly stated.} A sharp example, discovered recently, corresponds to Kerr BHs with scalar hair (KBHsSH)~\cite{Herdeiro:2014goa}. These are solutions for a free complex scalar field with mass $\mu$, minimally coupled to Einstein's gravity. They describe asymptotically flat, spinning BHs in equilibrium with scalar ``hair" around them. Both the geometry and the scalar field are regular on and outside the horizon, and the solutions interpolate continuously between (a subset of) vacuum Kerr BHs and horizonless gravitating solitons called \textit{boson stars}~\cite{Schunck:2003kk,Liebling:2012fv}. 
 
\begin{figure}[t!]
\begin{center}
\includegraphics[width=0.49\textwidth]{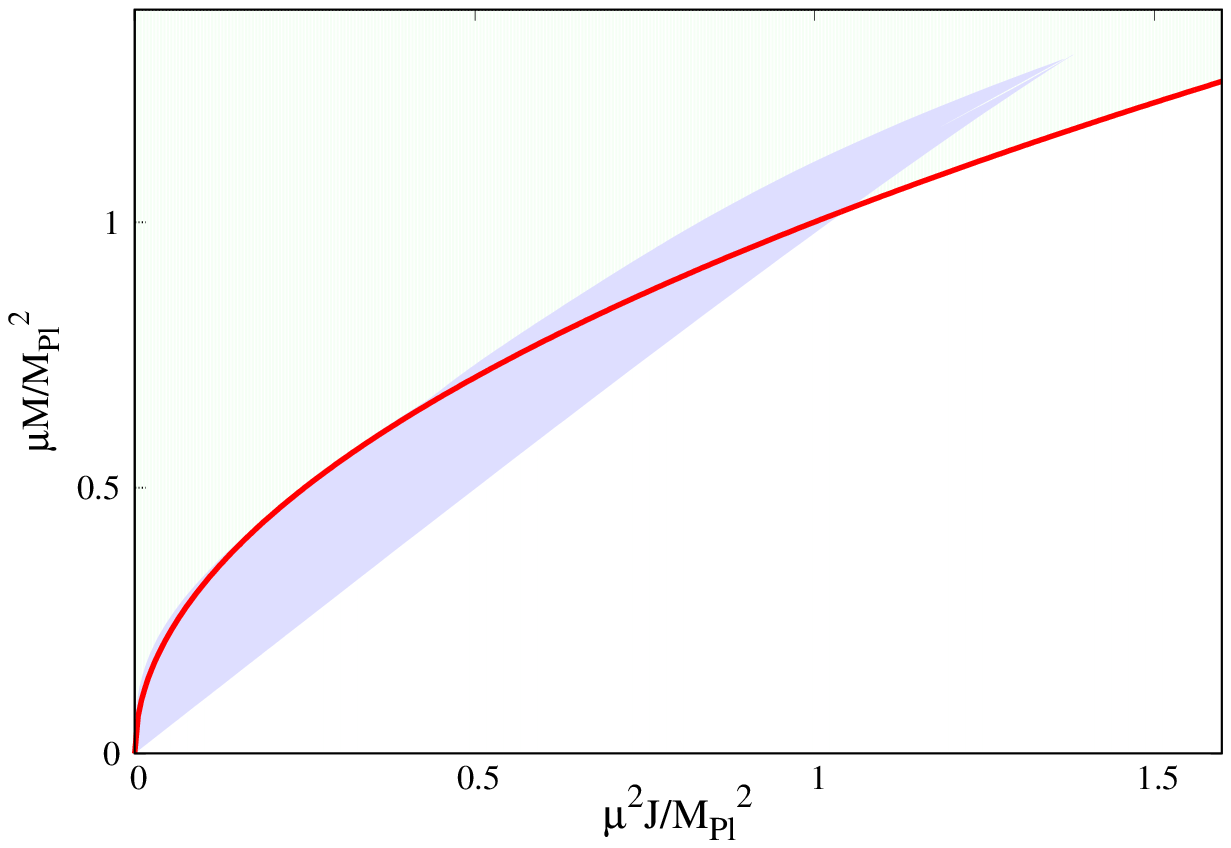}
\includegraphics[width=0.49\textwidth]{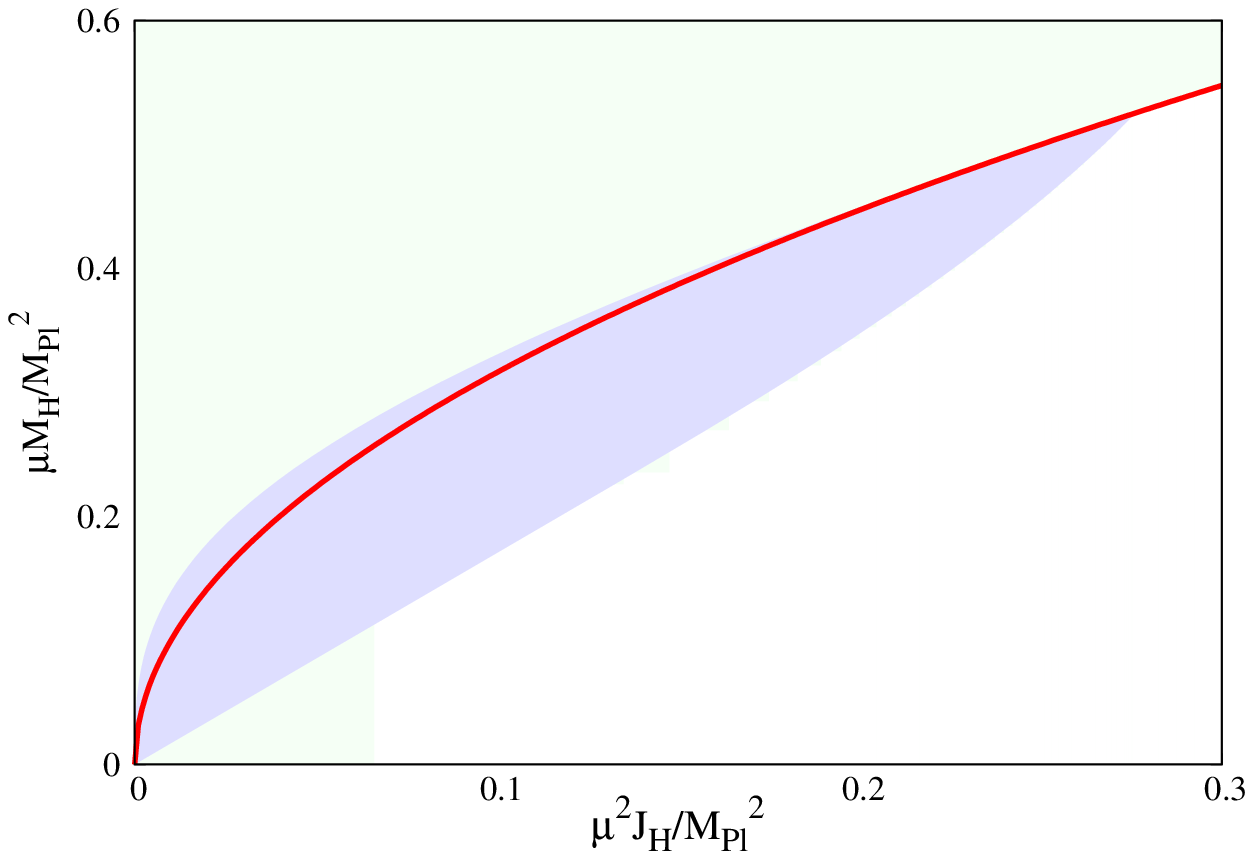}
\caption{\small{Space of solutions of KBHsSH, in terms of $M,J$ (left panel) and $M_H,J_H$ (right panel), 
corresponding to the blue shaded region, obtained from thousands of points, 
each point corresponding to one numerical KBHSH solution of the type described in~\cite{Herdeiro:2014goa}. The red solid line represents the Kerr bound and 
Kerr solutions exist only above this line, whereas KBHsSH exist also below.
$M_{Pl}$ is the Planck mass, $\hbar=1$ and we have used units set by 
the scalar field mass $\mu$, $cf$. \cite{Herdeiro:2014goa}.
}}
\label{fig1}
\end{center}
\end{figure}

In~\cite{Herdeiro:2014goa} it was observed that KBHsSH violate the Kerr bound - Fig.~\ref{fig1} (left panel), a result very recently independently confirmed~\cite{Kleihaus:2015iea}. This observation was made in terms of the ADM mass $M$ and angular momentum $J$; these take into account both the BH mass $M_H$ and angular momentum $J_H$
 -- which can be computed as Komar integrals on the horizon -- 
 as well as the mass, $M_\Psi$, and angular momentum, $J_\Psi$, 
 stored in the scalar field outside the horizon. It is possible for these BHs 
 to compute these individual contributions~\cite{Herdeiro:2015gia}: $M=M_H+M_\Psi$, $J=J_H+J_\Psi$, 
 and therefore it is legitimate to ask if the Kerr bound  violation remains in terms of the $M_H$, $J_H$, a question not yet addressed. This point is clarified in Fig.~\ref{fig1} (right panel), where it is shown that the Kerr bound is also violated in terms of horizon quantities for KBHsSH; indeed  
\begin{eqnarray}
 j_H\equiv \frac{cJ_H}{GM_H^2}
\end{eqnarray}
 takes values greater than one for some of the ``hairy" solutions.
 

\section{A different take: the Kerr bound and a velocity limit}
Having established that the \textit{strict} Kerr bound is not general, it is pertinent to ask if there is any universal bound on BH rotation. One bound was suggested in~\cite{Jaramillo:2011pg} for axi-symmetric, possibly dynamical and non-vacuum, spacetimes. It was formulated in terms of quasi-local quantities of an apparent horizon: its area $A_H$ and $J_H$. The bound reads $8\pi |J_H|\leqslant A_H$. We have checked that this bound is verified by the solutions described in Section 2. This bound, however, as the Kerr bound, is not truly a bound on BH \textit{rotation}, but rather on the BH's angular momentum, which is not the same. Indeed there are both rotations with zero total angular momentum -- for instance a cat malevolently dropped from rest upside down and which lands on its feet\footnote{Disclaimer: we have not performed this experiment ourselves.} --, and objects with total non-zero angular momentum which have no rotation. Both situations can be illustrated in BH physics, albeit higher dimensional, in the black Saturn system~\cite{Elvang:2007rd}. Here, we shall look for a bound on BH rotation, instead.  A detour through elementary physics is suggestive. 

\bigskip

In rigid body mechanics, the angular momentum for rotations around a principal axis of inertia is related to the angular velocity $\Omega$ by the moment of inertia $I$: $J=I\Omega$. A peripheral point, at distance $R$ from the axis has a linear velocity $v=RJ/I$. Imposing special relativity, $v\leqslant c$, one obtains a bound on the angular momentum $J\leqslant cI/R$. The moment of inertia is related to the total mass $M$ of the rigid body and the different scales of the object. Referring all of the latter to $R$ gives $I=\alpha MR^2$. The constant $\alpha$ is smaller than one if $R$ is the largest scale of the body, but it may be larger than one if larger scales exist. Thus $J\leqslant \alpha cMR$. Finally, imposing $R$ to be of the order of the gravitational radius leads, again, to an approximate Kerr bound: $j\lesssim \alpha$.

\bigskip

There are two suggestions from these heuristic arguments 
(even though a BH spacetime is \textit{not} a rigid body!). 
Firstly, that one may interpret a violation of the strict Kerr bound 
as due to a change in the rotational inertia of the system. 
This will be expanded in Section~6. 
Secondly, that an approximate Kerr bound also results from limiting linear velocities by $c$. This indicates it may be useful to define a \textit{horizon linear velocity} for rotating BHs.

  \newpage

\section{The horizon linear velocity and the $v_H$ bound}
We define a horizon linear velocity for asymptotically flat, stationary and axi-symmetric spacetimes as follows. Let  ${\bf m}$ be the Killing vector field associated with the $U(1)$ axi-symmetry. On a spatial section of the event horizon one computes the proper length of all closed orbits of ${\bf m}$. Let $L_{\rm max}$ be the maximum of all such proper lengths, which is finite, since the spatial sections of the horizon are compact and the orbits of ${\bf m}$ are closed. Then we define the \textit{circumferencial radius}, $R_c$, as 
\begin{equation} 
R_c\equiv \frac{L_{\rm max}}{2\pi} \ .
\end{equation} 
The horizon linear velocity, $v_H$, is defined as
\begin{equation}
v_H \equiv R_c \Omega_H \ ,
\end{equation}
where $\Omega_H$ is the angular velocity of the horizon, as usually defined by the Killing vector for which the  event horizon is a Killing horizon (see $e.g.$~\cite{Townsend:1997ku}). 


If one introduces coordinates adapted to the $U(1)$ symmetry, 
such that $m^\mu\partial_\mu=\partial/\partial \phi$, $\phi\in [0,2\pi[$, then $v_H$ is simply computed as
\begin{equation}
v_H= {\rm max}\left[\, \sqrt{g_{\phi\phi}}\Big|_{\rm horizon}\, \right] \Omega_H \ .
\end{equation}
For $\mathbb{Z}_2$ invariant BH solutions, $L_{\rm max}$ should occur at the $\mathbb{Z}_2$ invariant point, $i.e.$, on the equator. This is the typical case, but our definition allows for non-typical cases as well.

\bigskip

 We propose that for any four dimensional, stationary and axi-symmetric, asymptotically flat, rotating BH
\begin{equation}
v_H\leqslant c \ .
\label{vhbound}
\end{equation}

\section{Testing the universality of the $v_H$ bound}
\begin{figure}[h!]
\begin{center}
\includegraphics[width=0.48\textwidth]{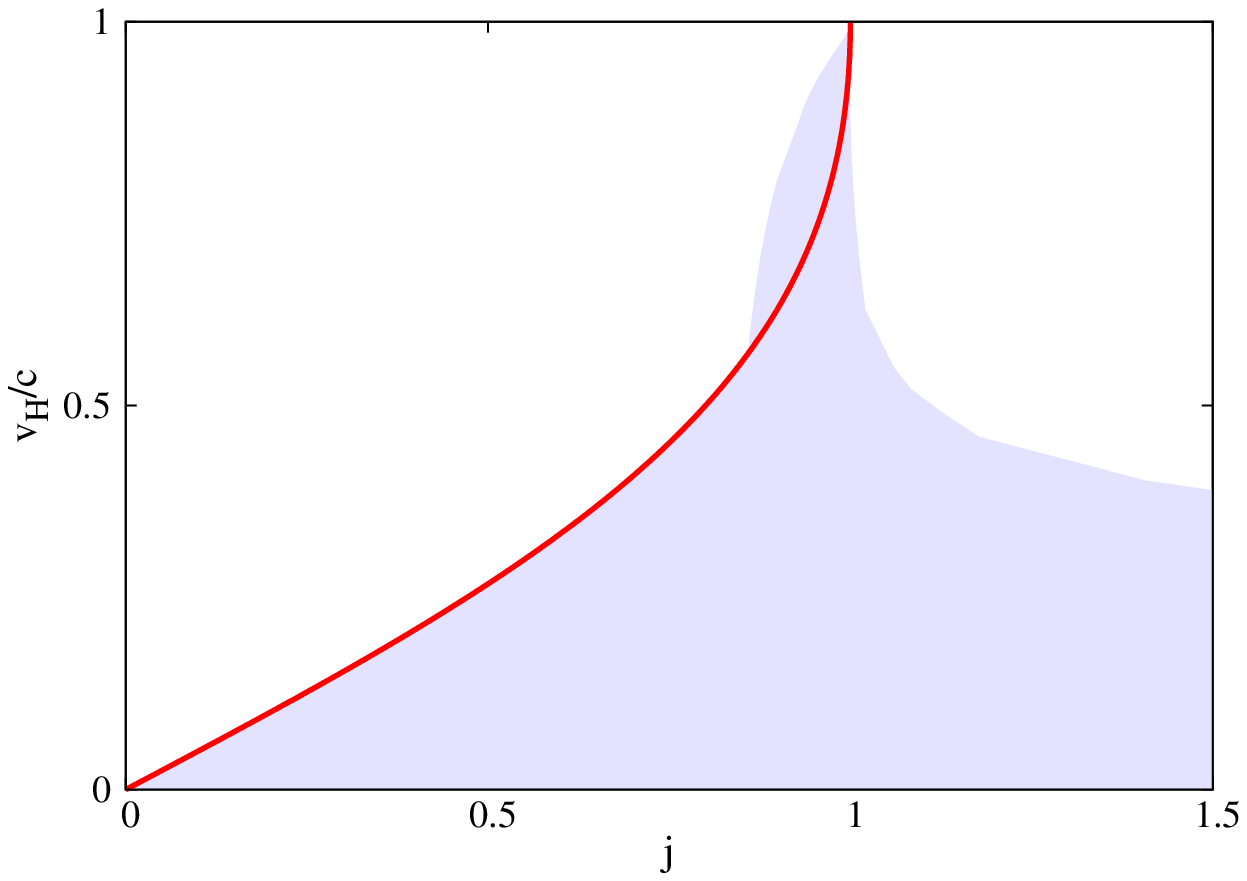}
\includegraphics[width=0.48\textwidth]{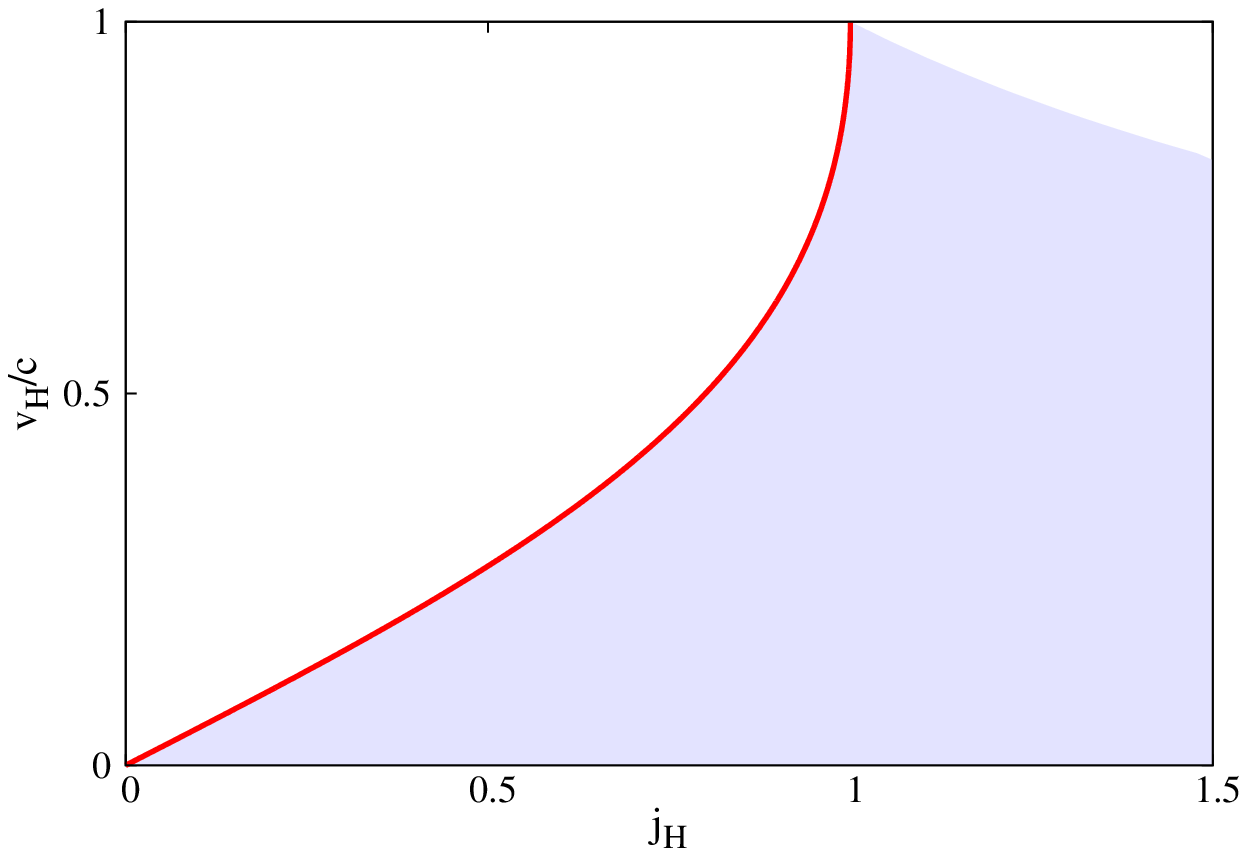}
\caption{\small{The horizon linear velocity $v_H$ $vs$. $j$ (left panel) or $j_H$ (right panel).}}
\label{fig2}
\end{center}
\end{figure}

To test the bound (\ref{vhbound}) we have plotted in Fig.~\ref{fig2} $v_H$ $vs.$ $j$ (left panel) and $v_H$ $vs.$ $j_H$ 
(right panel) for KBHsSH (blue shaded region) and also for the Kerr solution (red solid line), in which case
$
j=j_H={2v_H c}/{(c^2+v_H^2)} 
$.

The plots show that the $v_H$ bound is always verified, even when the Kerr bound is violated for KBHsSH. Moreover, the $v_H$ bound is only saturated for the extremal Kerr solution. 

 
 \bigskip
 
 We have also confirmed  (\ref{vhbound})  for 
the Kerr-Newman BH and other charged rotating solutions in general relativity.

\section{Discussion: the toll of heavy dragging\,?}

One interpretation for $v_H<c$ even when $j,j_H>1$ is the aforementioned variation in the rotational inertia. `Dirty' BHs, $i.e$ with surrounding matter, have additional ballast in their rotational motion, hence a larger ``moment of inertia". They are therefore able to accommodate more specific angular momentum than a vacuum Kerr BH, since for comparable horizon charges they have a lower horizon linear velocity. 
%
%
Such ``toll of heavy dragging" has been proposed in~\cite{Herdeiro:2009qy} based on the analysis of exotic solutions. KBHsSH are a clean example that supports the idea. Perhaps methods used in establishing bounds for quasi-local quantities, $e.g.$~\cite{Jaramillo:2011pg}, can be used to \textit{demonstrate} (\ref{vhbound}). 
%
%
 %
 %
 On the other hand, it has not escaped our attention that this bound does not apply straightforwardly to higher dimensional BHs. Generalizations that apply to this case will be discussed elsewhere.  
 
 \bigskip
 
 Finally, it is inspiring that, 100 years after general relativity was formulated, one is unveiling the basic postulate of \textit{special} relativity in answering the question posed as the title of this essay and concerning the (arguably) most fascinating object predicted by general relativity: \textit{the rotation of a BH cannot exceed the speed of light}.

\section*{Acknowledgements}
We would like to thank L. Lehner for a valuable discussion. The work of C.H. and
E.R. has been supported by the grants PTDC/FIS/116625/2010,  
NRHEP--295189-FP7-PEOPLE-2011-IRSES and by the CIDMA strategic funding UID/MAT/04106/2013.


\newpage



\end{document}